\documentclass[draft]{agujournal2019}
\usepackage{url} 
\usepackage{lineno}
\usepackage[inline]{trackchanges} 
\usepackage{soul}
\usepackage{amsmath}
\usepackage{float}
\draftfalse

\journalname{JGR: Space Physics}

\begin{document}

%
%

\title{Comparison of TIDI Line of Sight Winds with ICON-MIGHTI Measurements}

\authors{Chen Wu\affil{1}, Aaron J. Ridley\affil{1}}

\affiliation{1}{Climate and Space Sciences and Engineering, University of Michigan, Ann Arbor, Michigan, USA}

\correspondingauthor{Chen Wu}{chenwum@umich.edu}

\begin{keypoints}
\item Line of sight winds from four TIDI telescopes are compared to ICON-MIGHTI measurements
\item Performance figures of merit for individual TIDI telescopes in different satellite configurations are presented
\item Coldside TIDI telescopes match MIGHTI better than warmside ones in general, while systematic errors are found in Tel 1 during forward flight

\end{keypoints}

%
%

%
%


\begin{abstract}
The Thermosphere-Ionosphere-Mesosphere Energetics and Dynamics (TIMED) satellite has been making observations of the mesosphere and lower thermosphere (MLT) region for two decades. The TIMED Doppler Interferometer (TIDI) measures the neutral winds using four orthogonal telescopes. In this study, the line of sight (LOS) winds from individual telescopes are compared to the measurements from the Ionospheric Connection Explorer's (ICON's) Michelson Interferometer for Global High-resolution Thermospheric Imaging (MIGHTI) instrument from 90-100 km altitude during 2020. With the MIGHTI vector winds projected onto the LOS direction of each TIDI telescope, coincidences of the two datasets are found. The four telescopes perform differently and the performance depends on the satellite configuration and local solar zenith angle. Measurements from the coldside telescopes, Telescope 1 (Tel1) and Telescope 2 (Tel2), are better correlated with the MIGHTI winds in general with Tel2 having higher correlation coefficients across all conditions. The performance of Tel1 is comparable to that of Tel2 during backward flight while showing systematic errors larger than the average wind speeds during forward flight. The warmside LOS winds from Telescope 3 (Tel3) and Telescope 4 (Tel4) vary widely in magnitude, especially on the nightside. Compared with MIGHTI winds, the Tel4 measurements have the weakest correlation, while the Tel3 performance is comparable to that of the coldside telescopes during the ascending phase but deteriorates during the descending phase. Based on the TIDI/MIGHTI comparisons, figures of merit are generated to quantify the quality of measurements from individual telescopes in different configurations.
\end{abstract}

\section{Introduction}

Neutral winds play a crucial role in the thermosphere-ionosphere system. They are monitored with both ground- and satellite-based instruments. Typical ground-based instruments (meteor radar, medium-frequency radar, incoherent scatter radar, lidar, Fabry-Perot Interferometer (FPI), etc.) provide observations of neutral winds with a variety of temporal resolutions and coverages. These instruments are sparsely distributed
across fixed geophysical positions and provide nearly continuous data at a fixed latitude and longitude over a specific period. Satellites, on the other hand, orbit around the Earth, and over the course of a day, can provide global or near-global observations of the wind, allowing for studies of large-scale dynamics, depending on several assumptions. For example, the FPI \cite{hays1981} onboard the NASA Dynamics Explorer spacecraft mission measured the meridional winds primarily using the OI 630.0 nm emission that peaks at $\sim$240 km. It also measured wind profiles at 100-140 km with the 557.7 nm emission \cite{killeen1992}. On the same spacecraft, the Wind and Temperature Spectrometer (WATS) \cite{spencer1981} measured the in-situ zonal winds with the angle of arrival of the gas steam at altitudes from $\sim$300-700 km. The Upper Atmosphere Research Satellite (UARS) carried two instruments for neutral wind observations: the High Resolution Doppler Imager (HRDI) \cite{hays1993high} and the Wind Imaging Interferometer (WINDII) \cite{shepherd1993windii}. Both instruments measured Doppler shifts of the emission lines of photochemical species though using different techniques. HRDI provided measurements of horizontal winds at about 50-115 km during daytime and around 95 km during nighttime \cite{Burrage1994}; while WINDII measurements extended from 80 to 300 km. These earlier observations in the middle atmosphere deepened our understanding of the large-scale mesosphere and lower thermosphere (MLT) dynamics, especially revealing more details about migrating tides, nonmigrating tides, and planetary waves \cite<e.g.,>[and therein]{Morton1993,hays1994,forbes2003,wu1994, Killeen1988}.

The Thermosphere-Ionosphere-Mesosphere Energetics and Dynamics (TIMED) satellite was launched into a circular orbit at a nominal altitude of 625 km with an inclination of $\sim$74$^\circ$ in December 2001, with an aim to investigate and understand the energetics of the MLT region. The TIMED Doppler Interferometer (TIDI) on the spacecraft provides neutral wind measurements from 80-300 km altitude. It covers a wide latitude range that can extend to the northern or the southern pole depending on solar beta angle and season \cite{killeen1999timed,skinner2003,killeen2006timed,niciejewski2006timed}. Extensive studies have utilized the vector winds to investigate the characteristics of the migrating and nonmigrating tides, including the global distribution, propagation, long-term (monthly, seasonal, yearly, etc.) variations, and coupling with the ionosphere \cite<e.g.,>{wu2008_m,Wu2008_nonm,Oberheide2006,Oberheide2009,xu2009,oberheide2011_wave4,singh2018}. Also, due to the wide coverage of latitudes, more details of the planetary waves and gravity waves have been revealed \cite<e.g.,>{ortland2006,liu2009,chang2014quasi,gu2013,gu2021}. These waves dominating the MLT neutral winds are important for understanding the coupling between the upper and the lower atmospheres.

Comprising two decades of data collection in the MLT region, TIDI measurements are a valuable resource for the community to study the dynamics of the thermosphere-ionosphere system, especially the long-term variations. However, few thorough data validation studies have been conducted. This is partially due to the lack of proper datasets in the hard-to-measure MLT region. In October 2019, the Ionospheric Connection Explorer (ICON) was launched. The Michelson Interferometer for Global High-resolution Thermospheric Imaging (MIGHTI) onboard ICON provides a new wind dataset at low- and mid-latitudes from 90 to 300 km \cite{immel2018,englert2017michelson,harding2017,harding2021,makela2021}. Since the spacecraft's precession rate is more rapid than that of TIMED, measurement coincidences between TIDI and MIGHTI in time and space take place relatively frequently, providing a good opportunity to compare these two datasets. \citeA{dhadly2021} compared the vector winds of ICON-MIGHTI level 2.2 data and TIDI level 3 data which have an overlap between $\sim$90-120 km. Based on individual day comparisons, they found that the vector winds over the conjunctions of the two datasets are in agreement but the TIDI coldside measurements in forward flight show a systematic bias. The zonal and meridional winds from both products are calculated with the inverted altitude profiles of the line of sight (LOS) winds. Although the retrieving algorithms are different for TIDI and MIGHTI \cite{niciejewski2006timed,harding2017}, both include fittings and assumptions that possibly introduce “smoothing effects”.

In this study, LOS measurements from each of the four TIDI telescopes are investigated for details of the performance. For comparison purposes, the coincidences of measurements from TIDI and MIGHTI are investigated under different satellite configurations, i.e., forward/backward flight, ascending/descending phase, and different solar zenith angles (SZAs). Performance figures of merit for individual telescopes are generated to quantify the quality of TIDI LOS winds in comparison to the MIGHTI winds. 

\section{Data and Methodology}

\begin{figure}
    \noindent\includegraphics[width=15cm]{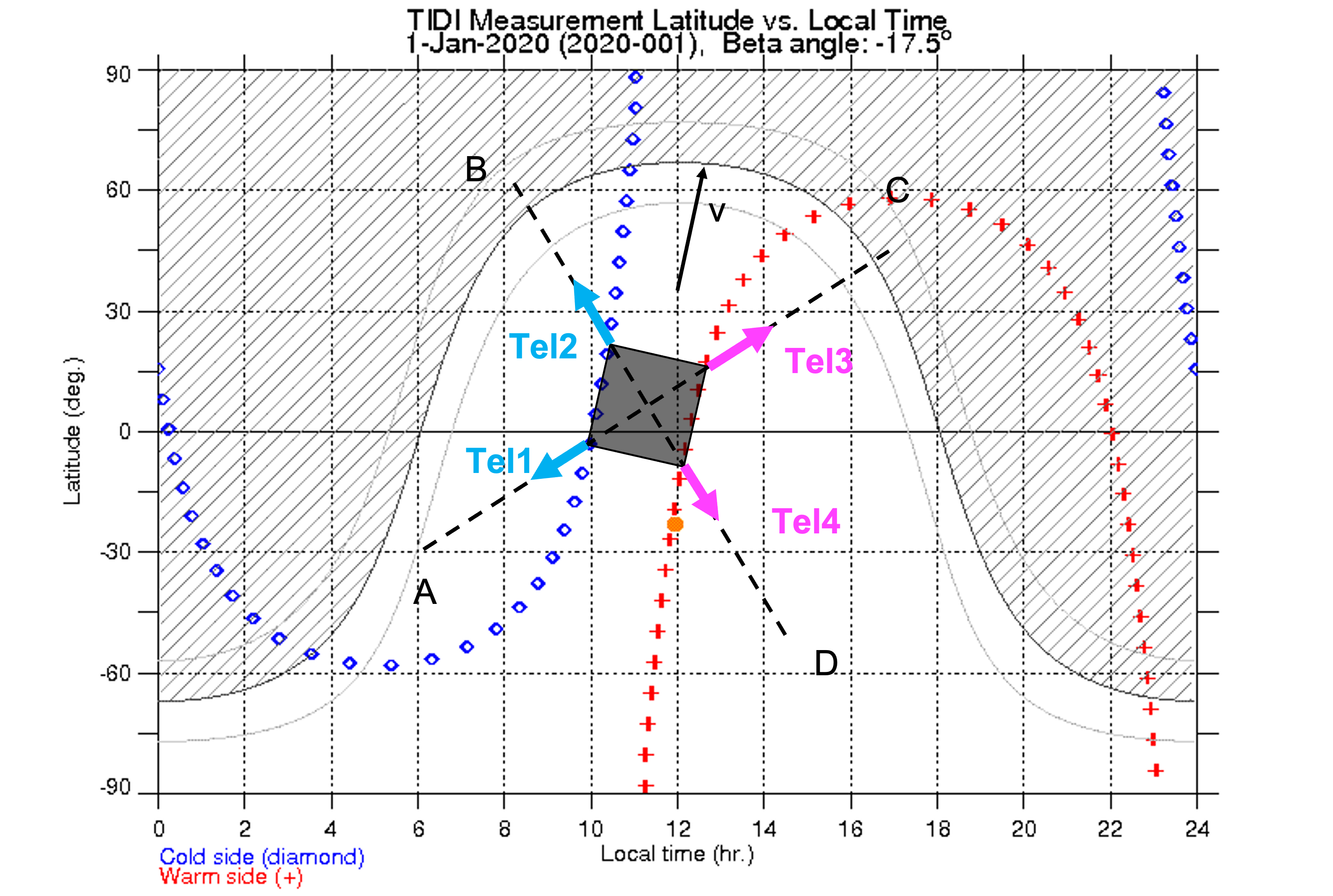}
    \caption{\raggedright Illustration of the geometry and measurements of TIDI based on an example on Jan 1, 2020.}
    \label{fig:geometry}
    \centering
\end{figure}

TIDI uses four orthogonal telescopes to observe the neutral temperature and winds simultaneously. The four telescopes are identical with $0.05^\circ\times2.5^\circ$ field of view scanning in directions $\pm45^\circ$ and $\pm135^\circ$ from the spacecraft velocity vector. Two of the telescopes are on the sunward side (warmside) and the other two are on the shadow side (coldside). Each side (each pair of telescopes) provides measurements from two local solar time (LST) sectors, one on each of the ascending and descending orbits. Thus measurements in four LSTs can be obtained every day. With a precession of $3^\circ/day$ (or $\frac{1}{5}$ of an hour of LST), it takes about 60 days for TIDI to cover all 24 LSTs if the ascending and descending orbits are combined. Figure \ref{fig:geometry} illustrates the geometry of the instrument and measurements on Jan 1, 2020, as an example. Telescope 1 (Tel1) and Telescope 2 (Tel2) are always on the coldside, while Telescope 3 (Tel3) and Telescope 4 (tel4) are always on the warmside. To keep the warmside and coldside telescopes always facing sunward and anti-sunward, respectively, TIMED makes a yaw maneuver changing from a forward flight configuration to a backward flight configuration or vice versa approximately every $\sim$60 days. The orbit was designed to make sure that the Earth-viewing geometry repeats every year, which means that TIDI monitors the same latitude at the same LST on the same day of the year \cite{killeen2006timed,niciejewski2006timed}. 

TIDI data products are provided at three levels \cite{niciejewski2006timed,killeen2006timed}: (1) Level 1 data product contains the raw spectra after the removal of instrumental and satellite-induced artifacts. The LOS brightness, background, and wind are derived with time and position annotated. (2) Level 2 product contains the inverted altitude profiles of winds on a uniform altitude grid. The inverted background and volume emission rate are also included. (3) Level 3 product provides zonal and meridional winds calculated with inverted LOS winds from level 2 data. In this study, the LOS winds from TIDI level 1 data product (Version 11) are investigated. There are two main processes to convert the TIDI data to level 1, as described by \citeA{niciejewski2006timed}, which is simply summarized as follows: 
\begin{enumerate}
    \item The raw spectra are checked for cosmic ray strikes, background corrected, and normalized with a white light calibration to remove all instrument and satellite-induced artifacts in the data.
    \item For each telescope, the processed spectra are fitted by a set of linear functions which are derived from the convolution of an instrument function and a normalized Gaussian line source function for each Doppler broadened line. Thus the LOS velocity, signal brightness, and continuum background are obtained. The LOS velocity at a given tangent altitude is a weighted average along the LOS direction of the velocity at each altitude. The “true” LOS wind is then calculated by shifting the “zero” wind position due to instrument temperature fluctuations, long-term instrument drift, the component of the spacecraft velocity along the LOS, and the component of Earth rotation along the LOS. 
\end{enumerate} The emission lines that are utilized to derive the level 1 LOS winds in the MLT region include O2 (0-0) P15 (765.07 nm), P9 (763.78 nm), broadband (764.00 nm), and OI 557.7 nm green lines. The measurements have an altitude coverage of $\sim$70-120 km ($\sim$70 - 120 km on the dayside and $\sim$80 - 105 km on the nightside) with an interval of $\sim$2.5 km in the MLT region. Details of the data can be found in the work by \citeA{niciejewski2006timed} and \citeA{killeen2006timed} and from the TIDI website (\url{http://tidi.engin.umich.edu/html/go?main.html&menu_home.html}).

MIGHTI is one of the four instruments on ICON, which was launched into a $27^\circ$ inclination by $\sim$575 km altitude orbit in October 2019. MIGHTI provides temperature and neutral winds measurements at low-/mid-latitudes from 90 to 300 km using two orthogonal Doppler Asymmetric Spatial Heterodyne interferometers \cite{immel2018,englert2017michelson}. For each sensor, the inverted LOS winds are derived based on the interferogram and recorded in the level 2.1 data product. Combining the inverted LOS winds in the two directions, the vector winds can be obtained, which are provided by level 2.2 data. For the green line (557.7 nm) emission measurements utilized in this study, the altitude range is $\sim$90-190 km during daytime and $\sim$90-109 km during nighttime with an altitude interval of $\sim$3 km. Due to the precession of the satellite, MIGHTI covers all 24 MLT hours at a given latitude in $\sim$27 days \cite{harding2017,harding2021,makela2021}.

\begin{figure}
\noindent\includegraphics[width=15cm]{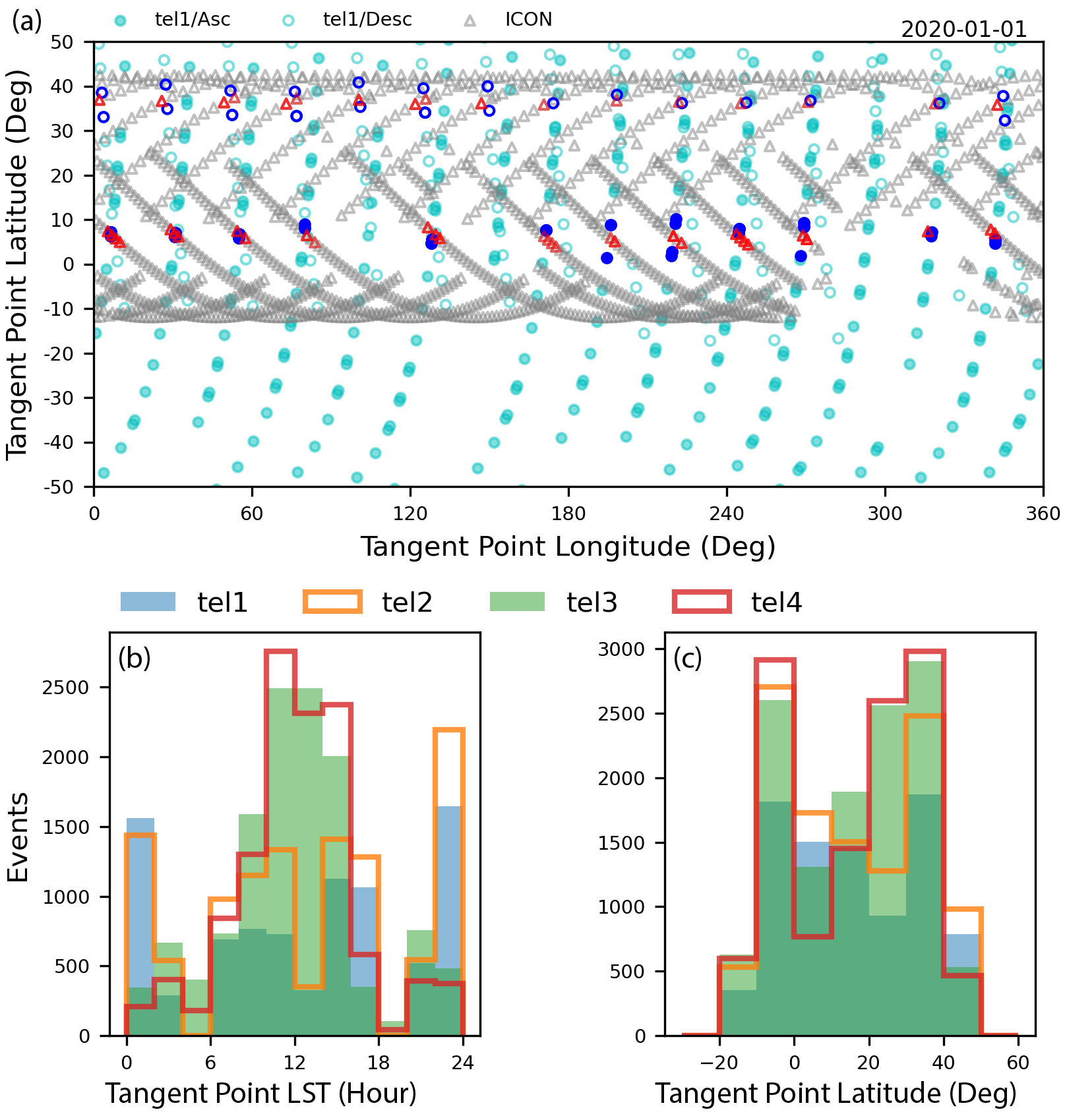}

\caption{\raggedright (a) Global map of locations of TIDI Tel1 LOS wind measurements (cyan filled circles for ascending and cyan empty circles for descending) and MIGHTI vector wind measurements (grey triangles) on Jan 1, 2020 at the altitude of $\sim$95 km. The red triangles represent the MIGHTI measurements falling within $\pm4^\circ$ latitude, $\pm4^\circ$ longitude, and $\pm$7.5 minutes of the TIDI measurements (blue filled and empty circles). The distributions of the LSTs (b) and the latitudes (c) at the tangent point of the coincident events for individual TIDI telescopes in 2020.}
\label{fig:hist}
\end{figure}

In this study, comparison of TIDI LOS winds with MIGHTI is based on coincidences of measurements from each TIDI telescope in 2020. The TIDI level 1 data and the MIGHTI level 2.2 data are used. Only the measurements with magnitudes larger than their uncertainty (i.e., signal-to-noise ratio (SNR) $>$ 1) were included for both instruments. Then the TIDI LOS winds were selected with (“data\_ok” = “True”) and (“p\_status” = 0). These are quality flags that are produced by daily processing routines. “data\_ok” is “True” meaning that the data is not contaminated; “p\_status” represents the processing status and no errors occur in deriving the LOS winds from raw spectra data if it is zero. Also, the measurements were limited with SZA less than 80$^\circ$ during daytime, SZA larger than 100$^\circ$ during nighttime, and solar scattering angle larger than 15$^\circ$ at the tangent point for optimum observations \cite{niciejewski2006timed}. For MIGHTI level 2.2 data, “ICON\_L22\_Wind\_Quality” labels the quality of wind measurements. We selected data with “ICON\_L22\_Wind\_Quality” equal to 1 which indicates the highest data quality. To find the coincidences, the LOS winds of TIDI were first binned with an interval of 2.5 km for each altitude profile. Figure \ref{fig:hist}a shows the measurements at 95 km from Tel1 on Jan 1, 2020, as an example. The MIGHTI observations were overplotted at similar altitudes. If one or more MIGHTI measurements existed in a spatial and temporal range of $\pm4^\circ$ latitude, $\pm4^\circ$ longitude, $\pm1.5$ km altitude, and $\pm$7.5 minutes around one TIDI data point, the MIGHTI observations were averaged and a coincident event was noted. To compare the LOS observations, the vector winds from MIGHTI were projected in the LOS direction of each TIDI telescope, which are termed “MIGHTI LOS” winds hereafter. Figures \ref{fig:hist}b and \ref{fig:hist}c show the distributions of the LST and the latitude at the tangent point for the coincident events for different TIDI telescopes in 2020. More coincidences took place on the dayside, especially for Tel3 and Tel4. For Tel1 and Tel2, there are a large number of events around midnight. Few events are found around the terminator due to the data selection criteria. Tangent point latitudes range from -20$^\circ$ to 50$^\circ$ with most events in the bins of $-10^\circ-0^\circ$ and $30^\circ-40^\circ$, due to ICON's orbit and viewing geometry. TIDI level 1 data provides MLT winds up to the altitude of $\sim$105 km on the nightside and $\sim$120 on the dayside. However, the measurement uncertainty increases significantly above 100 km and lots of measurements need to be discarded. In this study, measurements from 90-100 km are investigated and similar statistical results can be obtained with measurements larger than the uncertainty (i.e., SNR $>$ 1) from 100-120 km being included. Thus, the conclusions in this study can be applied to the measurements with SNR greater than 1 from 100-120 km.

\section{Results}

\subsection{TIDI/MIGHTI LOS Comparisons}

\begin{figure}
\noindent\includegraphics[width=15cm]{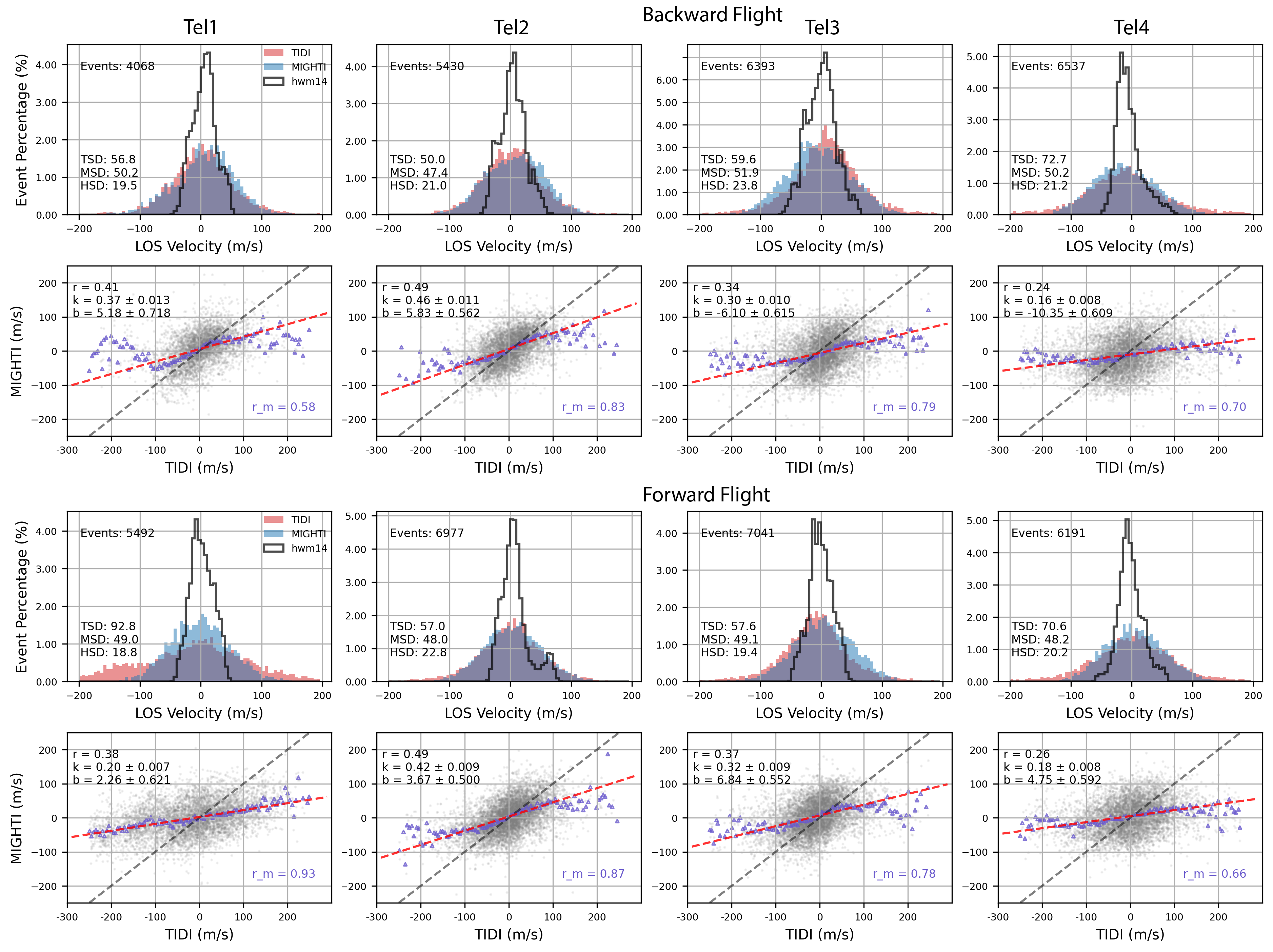}
\caption{\raggedright Comparisons of the TIDI and MIGHTI LOS winds from 90-100 km in 2020. The first and third rows are distributions of LOS winds for backward and forward flights, respectively; the standard deviation for HWM (HSD), TIDI (TSD), and MIGHTI (MSD) LOS winds are labeled. The second and fourth rows are point-to-point comparisons for backward and forward flights, respectively. The red dashed lines represent the linear least-squares regression fittings; the slope (k) and intercept (b) of the fitted line, as well as the correlation coefficient (r), are labeled in each subplot. The grey dashed lines represent y=x. The blue triangles denote the comparisons between TIDI velocity means and the corresponding MIGHTI averages. $r\_m$ are the correlation coefficients between the means.}
\label{fig:los_hist}
\end{figure}

Figure \ref{fig:los_hist} shows the distributions of TIDI and MIGHTI LOS winds in the 90-100 km altitude range during forward and backward flights for each telescope. Comparisons with the horizontal wind model (HWM) \cite{drob2014} are included. “HWM LOS” values were produced by projecting the zonal and meridional HWM winds at the TIDI measurement locations onto the LOS directions of individual telescopes. The distribution of HWM LOS winds looks very different from those of TIDI and MIGHTI LOS winds. The standard deviation (SD) of HWM (HSD) LOS is from $\sim$19 to $\sim$24 m/s, which is less than half of those of TIDI (TSD) and MIGHTI (MSD) LOS winds. Measurements from TIDI and MIGHTI are more consistent with each other than with HWM, with the exception of forward flight data which demonstrates longer distribution tails for TIDI Tel1 than MIGHTI. TIDI SD (approximately 50 to 60 m/s) is larger than that of MIGHTI by $\sim$10 m/s for Tel1/Backward, Tel2, and Tel3. Larger discrepancies are observed for Tel4 and Tel1/Forward (exceeding 70 and 90 m/s, respectively). \citeA{dhadly2021} cross-compared the MIGHTI and TIDI vector winds and found that the TIDI coldside measurements (i.e., from Tel1 and Tel2) in forward flight show a systematic bias. The distribution results indicate that this bias in the coldside vector winds is likely due to the Tel1 LOS winds with systematic errors, while the other coldside telescope, Tel2, does not show obvious average differences. Linear least-squares regressions were performed between the two datasets and the results are shown in the second and the fourth rows in Figure \ref{fig:los_hist}. Also, the TIDI LOS winds were binned with an interval of 5 m/s and the MIGHTI means were calculated in each TIDI velocity bin accordingly to show the comparisons of averages. Overall, Tel2 measurements are correlated best with the ICON LOS winds, while Tel4 observations show the worst correlation with the coefficients $<$ 0.3; Tel1 shows better results during backward flight than Tel3, whereas the slope of the Tel1 fitted line is smaller than that of Tel3 during forward flight configuration.

While the TIDI/MIGHTI comparisons have been described as coincident, it should be noted that these are gridded data, which can be anywhere in a volume of 8$^\circ$ longitude, 8$^\circ$ latitude, 3 km altitude, and 15 minutes. \citeA{harding19} noted that meridional winds between two FPI stations that are horizontally separated by $\sim$800 km could have cross correlations as low as $\sim$0.4. \citeA{larsen2002} noted that strong shears exist in the MLT region, which could significantly reduce correlations if there is any offset in altitude. Further, the comparisons that are performed here are between raw LOS winds from TIDI and inverted vertical wind profiles from MIGHTI.  It is expected that the raw LOS winds from TIDI are noisier than the inverted winds from MIGHTI. If there were no instrument noise at all, one may expect that the inverted winds would be closer to the truth winds, so would have more shear and would be “noisier”, and the raw wind measurements would be a brightness convolution of the true winds (i.e., a weighted summing), so would be more smooth than the inverted winds, but the instrument noise far outweighs the smoothing convolution, so the raw measurements (TIDI) are significantly more noisy than the inverted winds (MIGHTI).  Comparing the TIDI raw LOS data and the MIGHTI raw LOS data would be hard because the LOS measurements are not in the same direction. Comparing the TIDI inverted level 2 wind profiles with MIGHTI wind profiles makes sense, but adds another layer of possible injection of uncertainty into the TIDI data - if the inversion process is flawed in a systematic way, it would be difficult to know whether this was due to the data processing or the raw data.  Therefore, the significantly noisier TIDI raw LOS data was compared to the much smoother MIGHTI inverted winds.

Practically, this means that the TIDI winds have more noise, which impacts the cross correlation and the root mean square difference (RMSD) in the comparison to MIGHTI winds.  In combination with the geophysical offsets in location (reducing the correlation coefficient naturally), it is difficult to judge absolute point-to-point comparisons without a detailed investigation of every pass.  Instead, the focus is on relative comparisons, where the statistical comparisons were conducted for different telescopes in different configurations (SZA and flight direction), and the statistical differences were explored and are reported here.

\begin{figure}
\noindent\includegraphics[width=15cm]{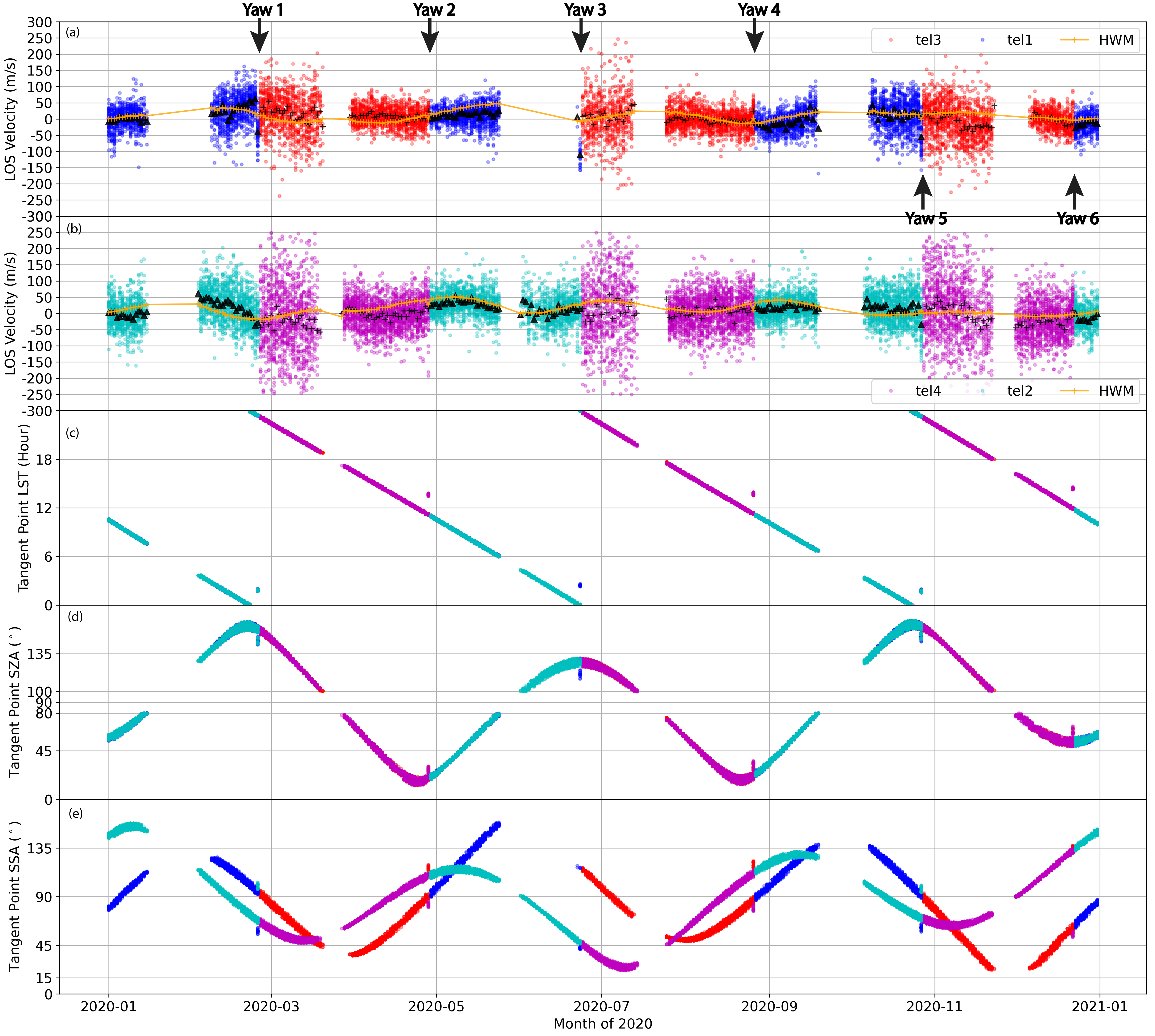}
\caption{\raggedright LOS winds in directions A (a) and B (b) that are denoted in Figure \ref{fig:geometry}, the LST (c), SZA (d), and SSA (e) as a function of the day of year at the latitude of $26^\circ$-34$^\circ$ and the altitude of 92-98 km in the ascending phase. The blue, cyan, red, and magenta dots represent Tel1 (coldside), Tel2 (coldside), Tel3 (warmside), and Tel4 (warmside), respectively. The black crosses and triangles represent the daily means of LOS winds from the warmside and coldside telescopes, respectively. Daily means of the HWM LOS winds are over-plotted as orange lines. The satellite is in backward flight on days 1-56, 119-175, 239-301, and 357-366, while it is in forward flight on days 56-119, 175-239, and 301-357.}
\label{fig:asc_tel12}
\end{figure}

\begin{figure}
\noindent\includegraphics[width=15cm]{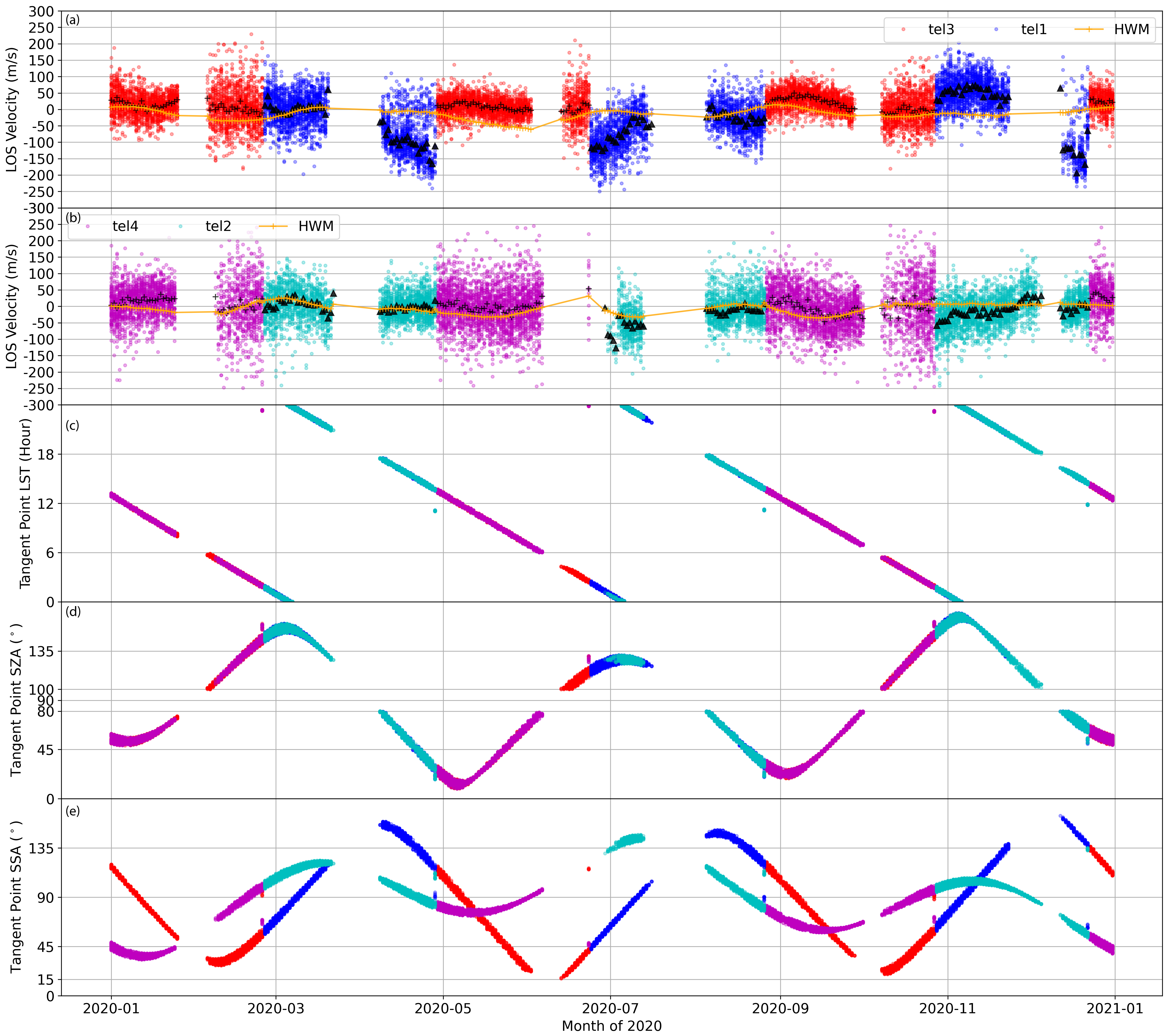}
\caption{\raggedright Similar to Figure \ref{fig:asc_tel12}, but for the measurements on the other side of the satellite, i.e., directions C and D in Figure \ref{fig:geometry}.}
\label{fig:asc_tel34}
\end{figure}

\begin{figure}
\noindent\includegraphics[width=15cm]{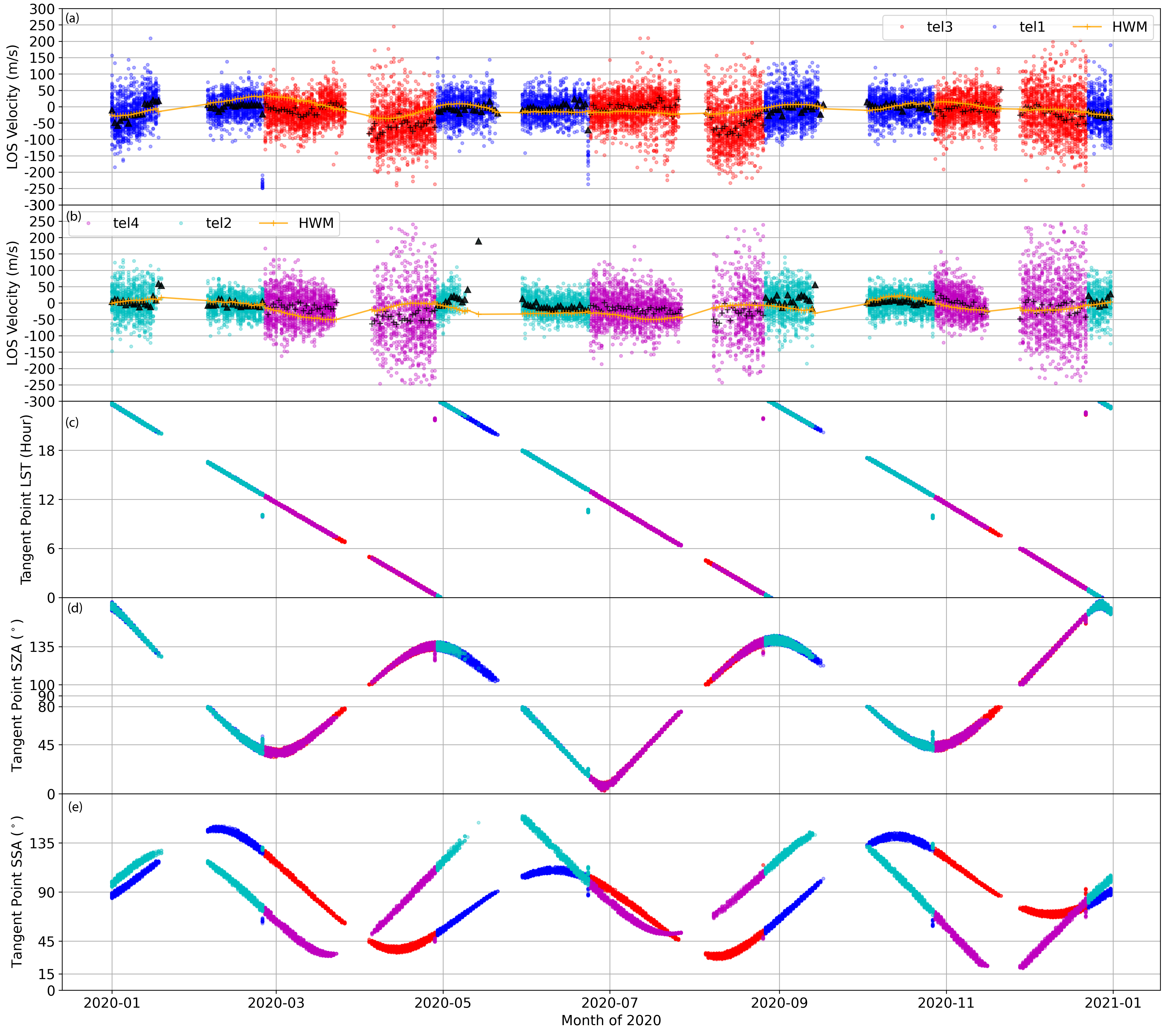}
\caption{\raggedright Similar to Figure \ref{fig:asc_tel12}, but for the descending phase.}
\label{fig:desc_tel12}
\end{figure}

\begin{figure}
\noindent\includegraphics[width=15cm]{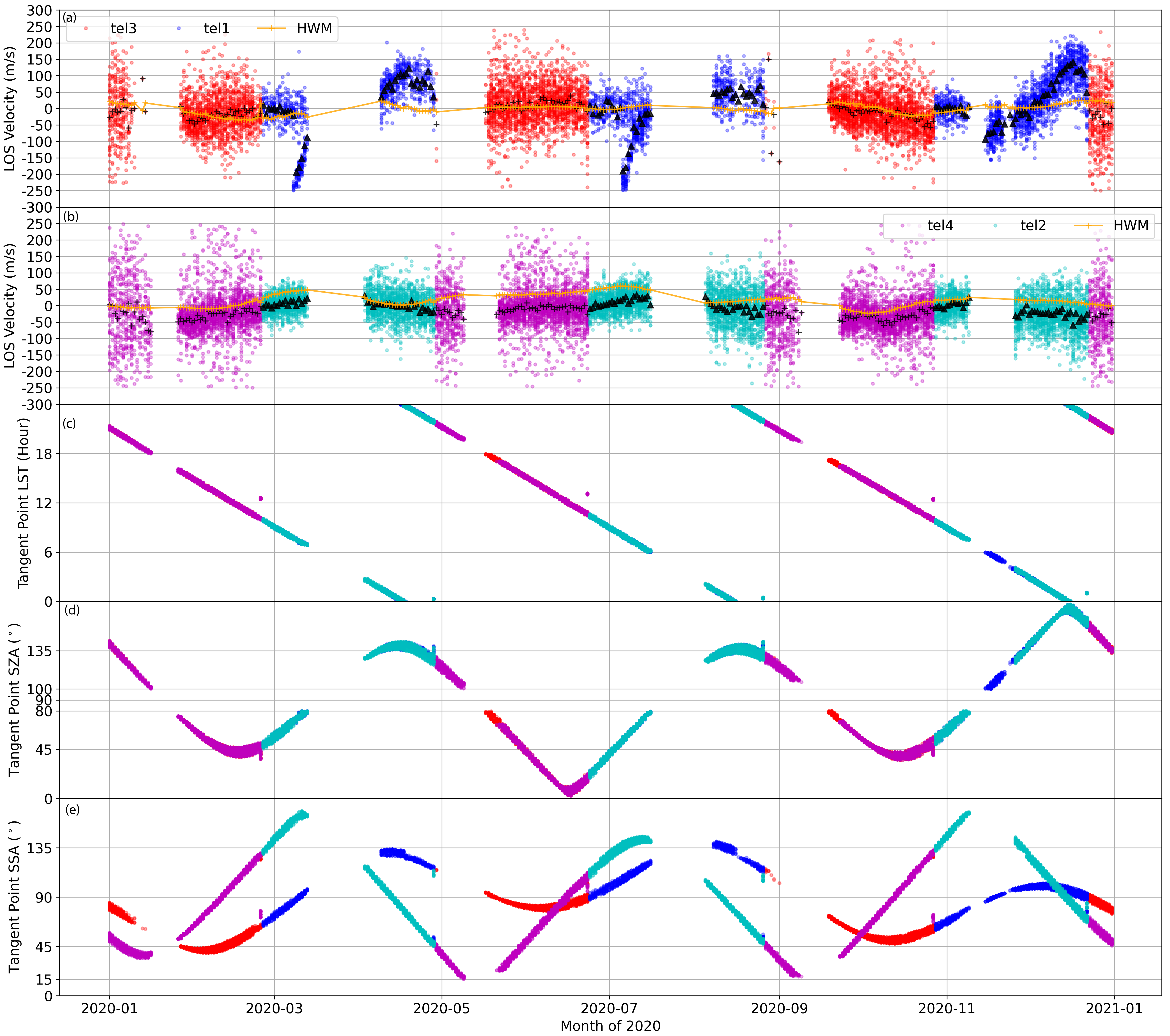}
\caption{\raggedright Similar to Figure \ref{fig:asc_tel34}, but for the descending phase.}
\label{fig:desc_tel34}
\end{figure}

Another way to think about the observations with TIDI is in a frame in which the four view directions (A-D) are fixed to the satellite velocity vector (through different LSTs) and the different telescopes contribute to the different look directions depending on whether TIMED is in forward or backward flight. Figure \ref{fig:asc_tel12} shows the LOS winds measured at the latitude of 26$^\circ$-34$^\circ$ from 92 to 98 km in directions A and B as a function of the day of the year in 2020 during the ascending phase. Each of the directions is monitored by a warmside telescope and a coldside telescope alternatively due to the yaw maneuver every $\sim$60 days. In 2020, the yaw maneuvers are made on days 56 (yaw 1), 119 (yaw 2), 175 (yaw 3), 239 (yaw 4), 301 (yaw 5), and 357 (yaw 6), as indicated. The satellite is in backward flight on days 1-56, 119-175, 239-301, and 357-366, while it is in a forward flight on days 56-119, 175-239, and 301-357. The daily means of the LOS winds were calculated and overplotted. The LST, SZA, and solar scattering angles (SSA) at the tangent point for each measurement were also plotted. The two telescopes on the same side, i.e., Tel1 (blue) and Tel2 (cyan) on the coldside or Tel3 (red) and Tel4 (magenta) on the warmside, have nearly the same LST and SZA (i.e., measuring the same location), but different solar scattering angles (i.e., looking in different directions). As a comparison, the daily means of HWM LOS winds were overplotted. Note that only TIDI LOS winds were investigated since there are not enough coincident measurements from MIGHTI to make this plot. Similar results for directions C and D are shown in Figure \ref{fig:asc_tel34}. The measurements in direction B (Figure \ref{fig:asc_tel12}b) are most consistent with the HWM means. In this direction, Tel2 (cyan) is in backward flight and Tel4 (magenta) is in forward flight. The daily means from these two telescopes are also consistent with each other. Unlike direction B, the TIDI LOS wind means deviate from the HWM means with differences of more than 100 m/s in direction C (Figure \ref{fig:asc_tel34}a). Systematic errors occur before and after the yaw maneuvers for Tel1/forward flight. In directions A (Figure \ref{fig:asc_tel12}a) and D (Figure \ref{fig:asc_tel34}b), the daily means are consistent each other and with HWM in general except for Tel2 measurements after yaw 5. Similar analysis was also conducted for the descending phase, as shown in Figures \ref{fig:desc_tel12} and \ref{fig:desc_tel34}. From the perspective of individual telescopes, the following conclusions are obtained:

\begin{enumerate}

\item{Tel1 has larger systematic errors (relative to HWM) during forward flight in both ascending and descending phases. The Tel1 means agree best with the HWM means during the descending/backward flight.}

\item{Tel2 only shows a slight systematic error in forward flight around November during the ascending phase. The measurements agree with the HWM means during most of the time but deviate by less than 50 m/s during the descending/forward flight.}

\item{Measurements from Tel3 and Tel4 are very scattered, especially on the nightside. The daily means for Tel3 generally agree with HWM means except for those during the ascending/backward flight. Tel4 daily means are consistent with HWM means in the ascending phase but have larger deviations in the descending phase.}

\end{enumerate}

\begin{figure}
\noindent\includegraphics[width=15cm]{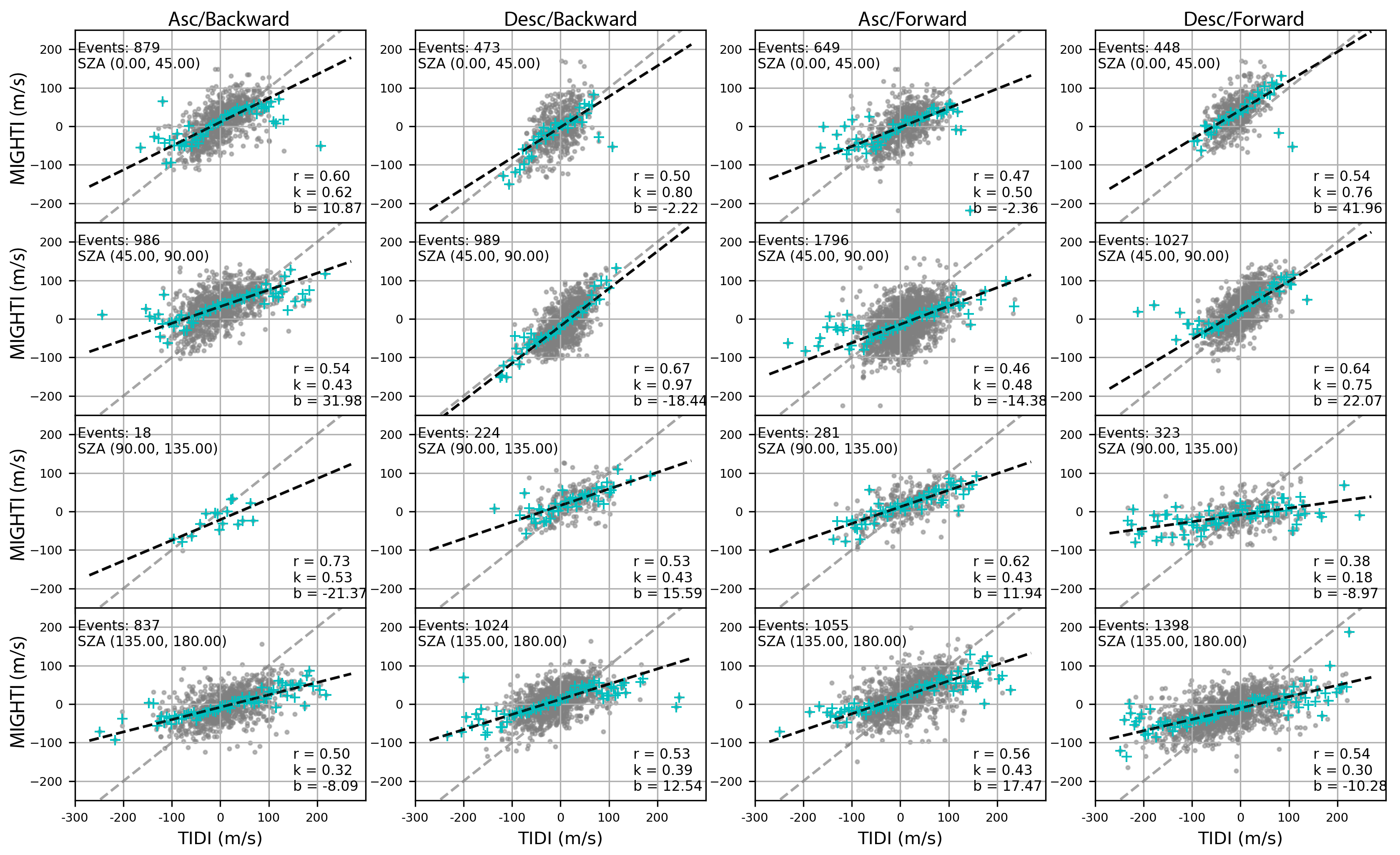}
\caption{\raggedright Point-to-point comparisons between TIDI Tel2 and MIGHTI LOS winds from 90-100 km in different SZA bins during Asc/Backward (first column), Desc/Backward (second column), Asc/Forward (third column), and Desc/Forward (fourth column) flights. The black dashed lines represent the linear least-squares regression fittings; the slope (k) and intercept (b) of the fitted line, as well as the correlation coefficient (r), are labeled in each subplot. The cyan crosses denote the comparisons between TIDI velocity means and the corresponding MIGHTI averages. The grey dashed lines represent y=x.}
\label{fig:sza_tel2}
\end{figure}

\begin{figure}
\noindent\includegraphics[width=15cm]{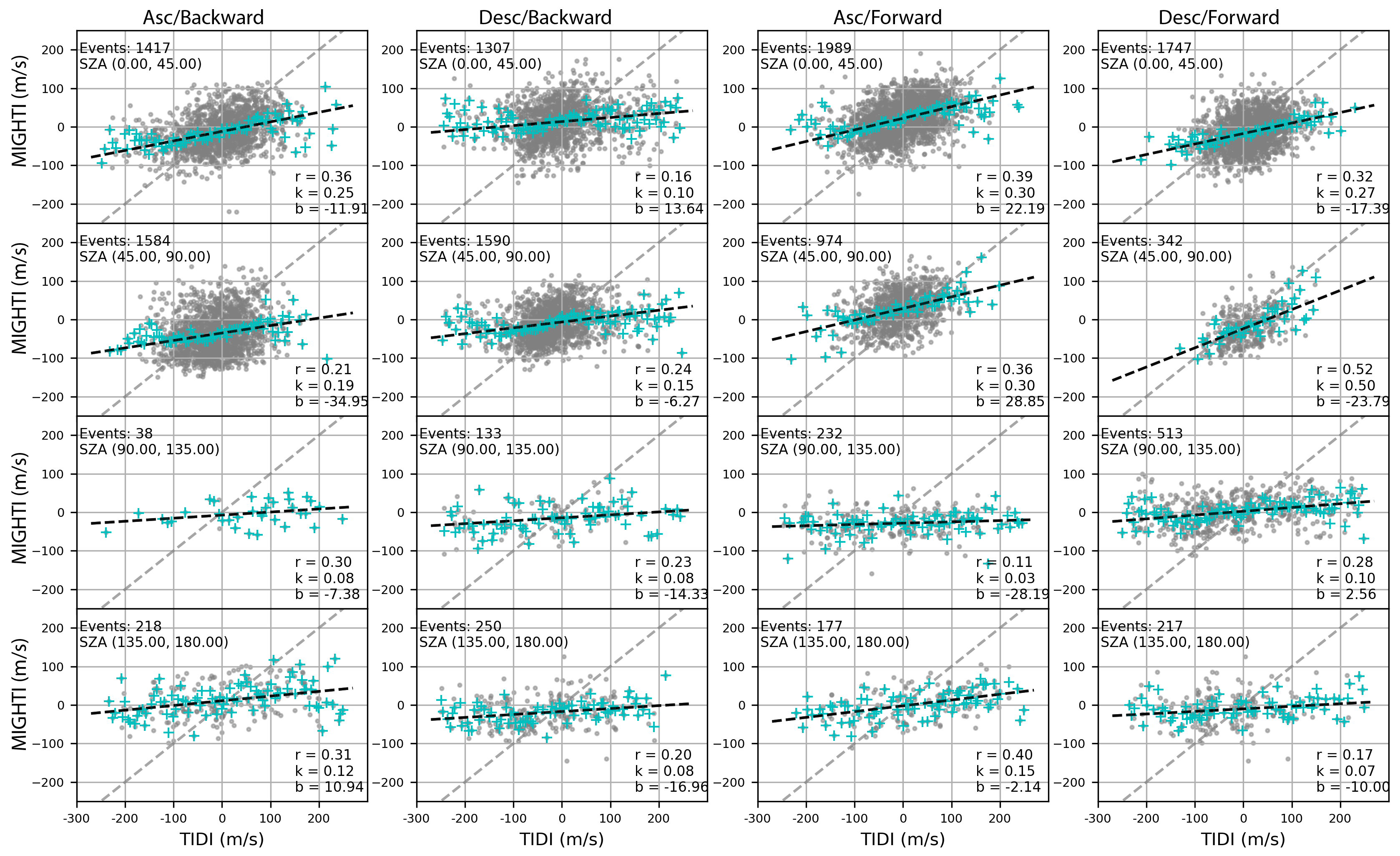}
\caption{\raggedright Similar to Figure \ref{fig:sza_tel2}, but for Tel4.}
\label{fig:sza_tel4}
\end{figure}

To further investigate the performance of the telescopes, the TIDI/MIGHTI coincidences were organized from $0^\circ-180^\circ$ SZA with an interval of $45^\circ$ in different satellite configurations: ascending/backward flight (Asc/Backward), ascending/forward flight (Asc/Forward), descending/backward flight (Desc/Backward), and descending/forward flight (Desc/Forward). In each SZA bin, the linear least-squares regression fitting was performed between the TIDI and MIGHTI LOS winds. Figures \ref{fig:sza_tel2} and \ref{fig:sza_tel4} are the results for Tel2 and Tel4 which show the best and the worst correlation with MIGHTI, respectively. The TIDI LOS winds were binned with an interval of 5 m/s and the MIGHTI means were calculated in each velocity bin accordingly to show the comparisons of averages. Tel2 and MIGHTI have a correlation coefficient larger than $\sim$0.5 for most of the SZA bins with better correlation on the dayside than on the nightside. The correlation coefficients between Tel4 and MIGHTI are less than $\sim$0.4 in most SZA bins. The small magnitudes of slopes of the fitted lines also indicate inconsistency between the two datasets, especially on the nightside.

\subsection{Performance Figures of Merit}

The TIDI/MIGHTI comparisons indicate that the four telescopes perform differently and the performance depends on satellite configuration (Asc/Backward, Desc/Backward, Asc/Forward, and Desc/Forward) and SZA (LST). To quantify the quality of the measurements from individual telescopes, figures of merit are generated.

\begin{figure}
\noindent\includegraphics[width=15cm]{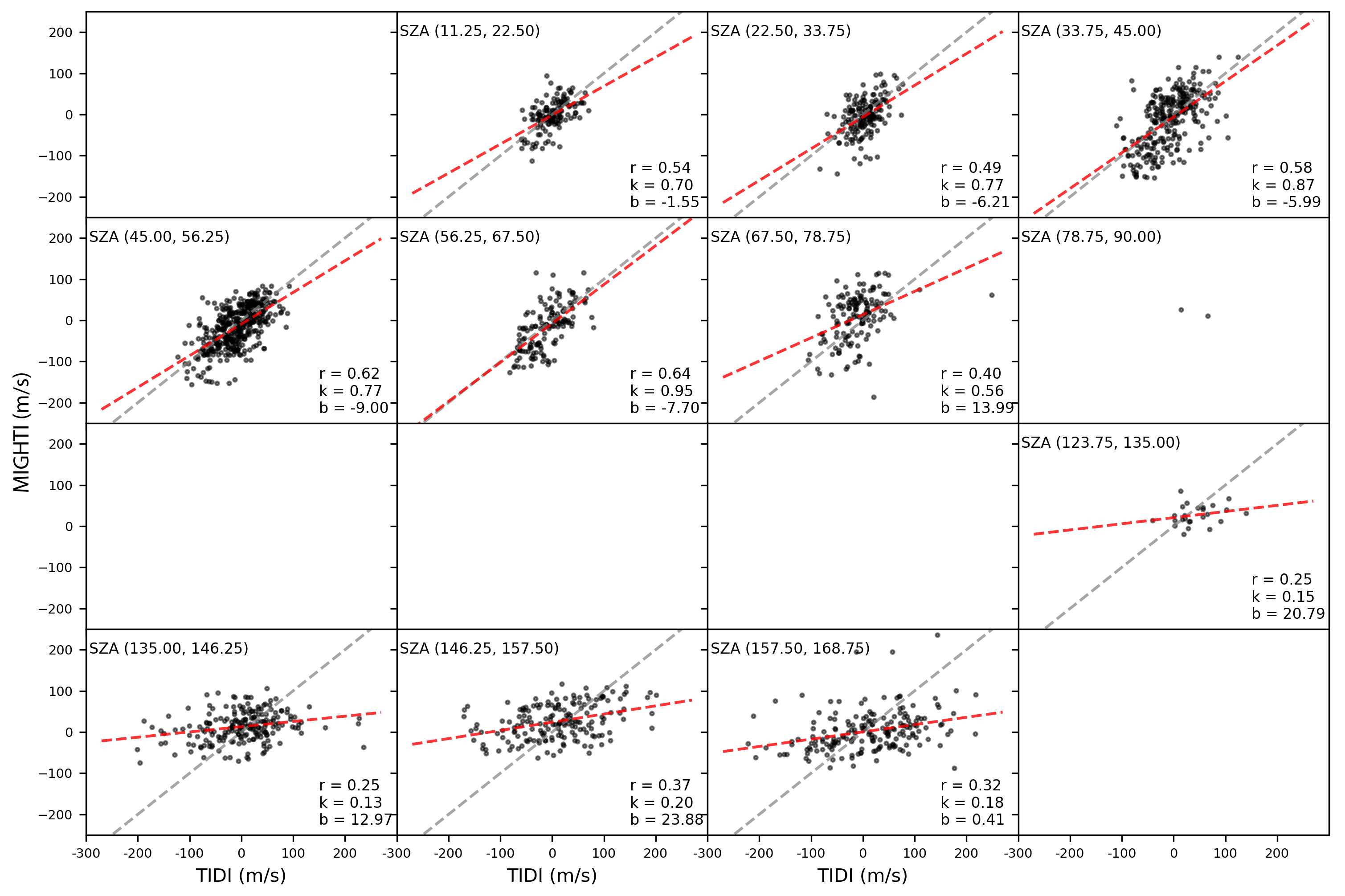}
\caption{\raggedright Point-to-point comparisons between TIDI Tel1 and MIGHTI LOS winds from 90-100 km in different SZA bins during Asc/Backward flight. The red dashed lines represent linear least-squares regression fittings; the slope (k) and intercept (b) of the fitted line, as well as the correlation coefficient (r), are labeled for each SZA bin. The grey dashed lines represent y=x.}
\label{fig:tel1_sza_merit}
\end{figure}

The coincidences of TIDI and MIGHTI LOS winds were re-organized by the SZA bins for each satellite configuration and the analysis of the correlations between the two datasets was conducted for each SZA bin, which is similar to Figures \ref{fig:sza_tel2} and \ref{fig:sza_tel4} but the SZA bin range is 11.25$^\circ$ instead of 45$^\circ$. Figure \ref{fig:tel1_sza_merit} shows an example during Asc/Backward flight for Tel1. Ideally, the slope (k), intercept (b), and correlation coefficient (r) should be equal to 1, 0, and 1, respectively. The closer the parameters are to the ideal values, the higher the consistency between datasets. Thus a score ranging from 0 to 10 was calculated for each parameter: 0 for completely inconsistent, 10 for completely consistent, with a linearly scaled score in between. Table 1 shows the cutoffs used to define the lowest and highest scores for each parameter. The total score in each SZA bin was then the mean of the three scores. To evaluate the dayside measurements, a weighted average of the scores was calculated from 0$^\circ$ to $90^\circ$ SZA with the weights determined by the total numbers of events in different SZA bins. The nightside score was calculated similarly but for $90^\circ-180^\circ$ SZA.

\begin{table}
    \caption{Score cutoffs for slope (k), intercept (b), and correlation coefficients (r)}
    \centering
        \begin{tabular}{l c c c}
        \hline
        Score  &Slope &Intercept (m/s) &Corr  \\
        \hline
        $0$      &$k<0.1$   &$|b|>50$   &$r<0.2$ \\
        $10$       &$|k-1|<0.1$   &$|b|=0$   &$r>0.9$ \\
        \hline
        \multicolumn{2}{l}{$^{a}$Scores are linearly scaled between the cutoffs.}
     \end{tabular}
     \label{table:cutoff}
\end{table}

\begin{figure}
\noindent\includegraphics[width=15cm]{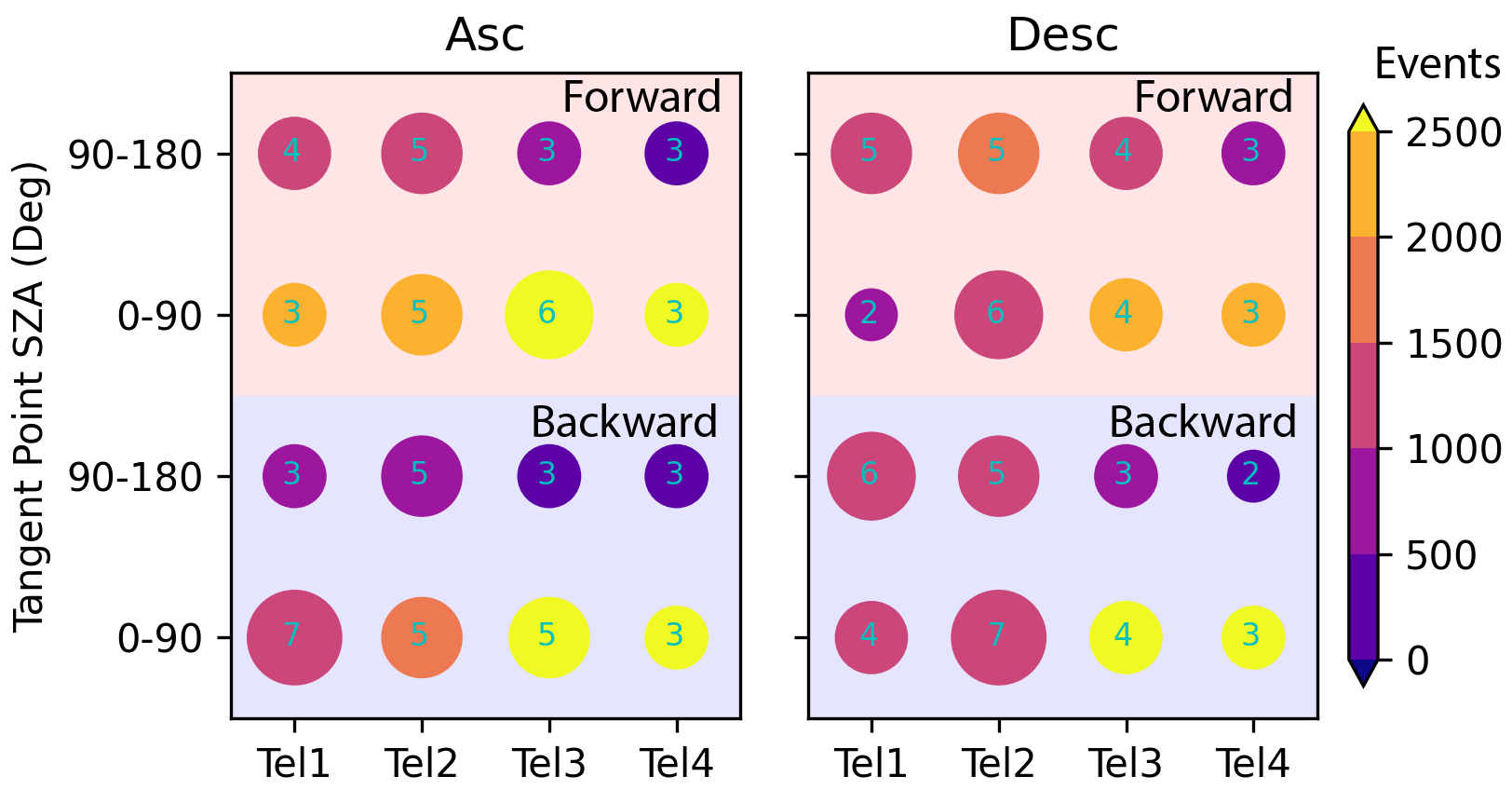}
\caption{\raggedright Performance figures of merit for each telescope on the dayside ($0^\circ-90^\circ$ SZA) and the nightside ($90^\circ-180^\circ$ SZA) during the ascending (left) and descending phases (right). The blue and red background colors represent the backward and forward flights, respectively. In each configuration, the score is represented by the size of the circle and labeled; the color of the circle represents the number of events.}
\label{fig:merits}
\end{figure}

Figure \ref{fig:merits} shows the performance figures of merit for the four telescopes in each of the satellite configurations. The measurements by Tel2 show the best results in general, especially on the dayside during the descending phase, while Tel4 performs the worst overall with all scores less than 4. Tel3 is comparable with Tel2 on the dayside during the ascending phase, whereas it deteriorates during the descending phase and on the nightside. The low scores for Tel1 in forward flight are due to the systematic errors as discussed above. During backward flight, Tel1 performs better than the warmside telescopes and is comparable to Tel2 except for the nighside in the ascending node. This is consistent with the results above. Additionally, the RMSD was calculated for each satellite configuration and telescope. Basically, with a score $<$ 5, the RMSD between the two datasets is $\sim$60-110 m/s; for a score $>=$ 5; the RMSD between the two datasets is $\sim$50-60 m/s. Thus, the RMSDs and the figures of merit are consistent.

It should be noted that the figure of merits that are presented here are not meant to substitute for a quantitative uncertainty analysis for any specific measurement point. They are indications of general quality of the data and should be treated as a quality flag. For example, if the figure of merit is 3 or less in a telescope/flight configuration, the general quality can be thought of as “poor”, while if the figure of merit is 5 or higher, the general data quality can be thought of as “good”. With further analysis, it is hoped that this quality flag can be refined to a quantified uncertainty in the data.

\section{Discussion and Summary}
The TIDI LOS winds provided by the level 1 product (Version 11) are compared to the MIGHTI observations from 90-100 km altitude during 2020. The zonal and meridional winds from the ICON-MIGHTI level 2.2 data are projected in the LOS direction of each TIDI telescope, which is termed “MIGHTI LOS” winds in this study. The coincidences of TIDI and MIGHTI LOS winds are mainly distributed from $\sim$-20$^\circ$ to $\sim$50$^\circ$ latitude with more events on the dayside than on the nightside.

The performance of individual telescopes is different and varies depending on satellite configuration and SZA/LST, with the coldside telescopes generally performing better than the warmside ones. TIDI Tel2 measurements correlate best with the MIGHTI LOS winds; Tel1 performance is comparable to that of Tel2 during backward flight except for Asc/Night, but has systematic errors during forward flight. Both of the warmside telescopes have more scattered LOS winds, especially on the nightside for Tel4. Tel4 shows the worst comparisons to the MIGHTI LOS winds, although no obvious systematic errors are found. Tel3 performs as well as the coldside telescopes on the dayside during the ascending phase but deteriorates during the descending phase and on the nightside. In terms of systematic errors in the measurements, only Tel1 LOS winds demonstrate significant systematic errors larger than the average wind speeds during forward flight. Results in this study agree with work by \citeA{dhadly2021} reporting a systematic bias in the coldside vector winds from TIDI during forward flight and further indicate that the reported coldside bias is very likely from Tel1. Similar plots to Figures \ref{fig:asc_tel12}-\ref{fig:desc_tel34} were made for measurements in previous years (not shown). The systematic errors in Tel1 during forward flight began to occur in 2015.

To quantify the quality of the data, the performance figures of merit for individual telescopes are generated based on the correlations and the linear regression fittings between TIDI and MIGHTI LOS winds. Figure \ref{fig:merits} shows that Tel2 and Tel4 are correlated the best and the worst with MIGHTI, respectively. Tel1 is slightly better than Tel3 during backward flight. During forward flight, Tel1 has a score of $\sim$2 on the dayside suggesting the worst comparison, while the Tel1 nightside score indicates better performance than that of Tel3. 

\begin{figure}
    \noindent\includegraphics[width=15cm]{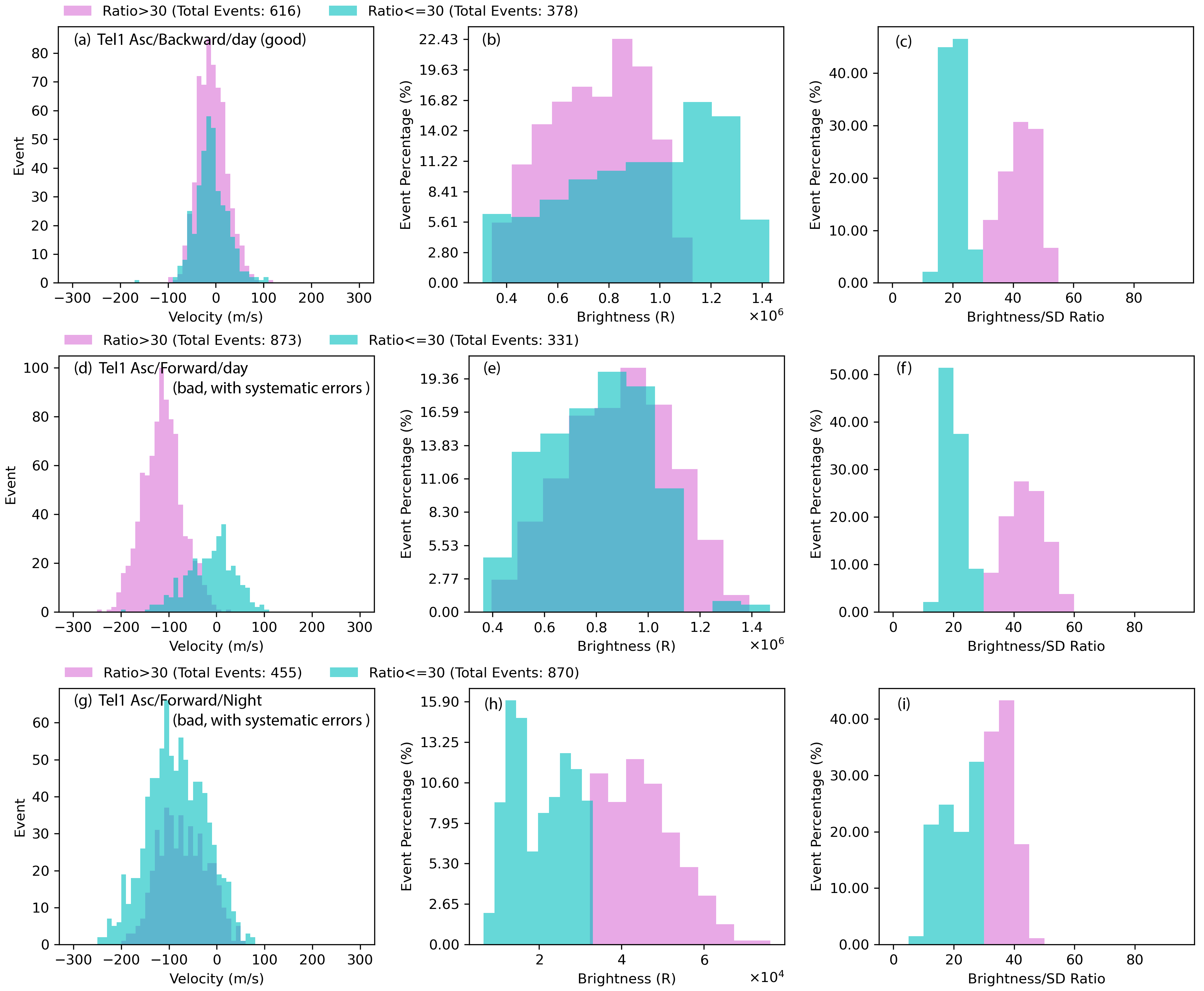}
    \caption{\raggedright LOS wind distributions(left), brightness percentage distributions (mid), and brightness ratio percentage distributions (right). From the top to the bottom are three examples of Tel1 measurements matching well with MIGHTI (a-c, on the dayside during Asc/Backward flight), with systematic errors and two different velocity distributions (e-f, on the dayside during Asc/Forward flight), and with systematic errors and two similar velocity distributions (on the nightside during Asc/Forward flight), respectively. Measurements with brightness ratios $<=$ 30 and $>$ 30 are colored with cyan and pink, respectively.}
    \label{fig:brightness}
    \centering
\end{figure}

As mentioned in the section of data and methodology, the LOS winds, brightness, and their variances are determined from the calibrated spectra by fitting a set of linear functions which are derived from the convolution of an instrument function and a normalized Gaussian line source function for each Doppler broadened line. The uncertainty of the horizontal winds along the LOS depends on the signal brightness and the suitability of the Gaussian fitting function \cite{niciejewski2006timed}. Therefore, to better understand the possible causes of the poor performance in Tel1, the brightness, the brightness SD, and the ratio of the brightness to the brightness SD from Tel1 during $\sim$30 days before or after each yaw maneuver in forward flight in 2020 were investigated. The brightness SD is the square root of the brightness variance which is the uncertainty in the brightness estimation. The systematic errors occur around yaws 2, 3, 4, 5, and 6 during the ascending phase (Figure \ref{fig:asc_tel34}a) and yaws 1, 2, 3, 5, and 6 during the descending phase (Figure \ref{fig:desc_tel34}a). Among these 10 periods, there are 5 periods where Tel1 measurements have two different velocity distributions depending on brightness ratio. An example for days 89-119 (before yaw 2) during the ascending phase is shown in Figures \ref{fig:brightness}d-f. For observations with the ratio $<=30$ (cyan), the LOS velocities are distributed from -150 to 100 m/s peaking around zero, while the velocity distribution with ratio $>$ 30 (pink) ranges from -250 to 0 m/s peaking around -100 m/s. The latter group is associated with the systematic error identified in Figure \ref{fig:asc_tel34}. In contrast, for measurements with good quality, the velocities of the two ratio groups are similarly distributed and peak around zero. An example for days 209-239 (after yaw 4) during the ascending phase is shown in Figure \ref{fig:brightness}a-c. Consequently, in order to reduce the systematic errors, the Tel1/Forward measurements on the dayside (i.e., Figures \ref{fig:brightness}d-f) with ratio $>$ 30 could be discarded. However, it should be noted that this is a rough investigation to see if the Tel1/Forward systematic errors can be restricted by the signal brightness. There are other scenarios where the measurements with systematic errors cannot be discarded with the ratio restriction discussed above. For example, Figures \ref{fig:brightness}g-i show the distributions of the nightside measurement on days 175-205 from Tel1. Both the measurements with the ratio $>$ 30 and $<=$ 30 are negatively biased. Therefore, although the brightness that is related to the derivation of LOS winds can be applied to exclude part of the “bad” measurements, it cannot be used to locate all for the Tel1/Forward winds. And we do not know the cause of the double-peaked distribution of brightness ratio at this time, but it is an avenue we are pursuing to get to the bottom of the data quality issues.

There are two known anomalies of TIDI after the launch of the satellite, which has been discussed by \citeA{killeen2006timed} and \citeA{skinner2003}: (1) the high and variable background of the instrument signal during daytime observation that was caused by a light leak of the profiler, and (2) the broadening of the spectral distribution and the increase of cross-talk between different telescope fields that was caused by ice formation on the detector housing window or optical surfaces. The TIMED satellite made two roll maneuvers in early 2003 to increase the detector housing temperature and thus most of the ice was removed. Also, measurements of the O2 (0-0) P branch broadband emission were favored to increase the signal level, since the background contamination was not as strong in this band, so the effective SNR was increased. Although these effects have been accounted for in the data processing, they are still possible sources of measurement noise and differences between TIDI and MIGHTI. Further, the TIDI instrument was designed to map the five fields (four telescope and one calibration) on the detector in the order of: calibration, Tel1, Tel2, Tel3, and Tel4 away from the center. The background intensity increases away from the vertex, indicating a lower signal-to-noise level in the outer field where the warmside telescope fields are imaged. This may contribute to the poor performance of warmside telescopes which have a larger scatter in their LOS winds. In addition, the estimation of a “zero wind” is a major source of uncertainty in neutral wind measurements for all orbiting FPI instruments, which may also contribute to the differences. For TIDI data processing, the “zero wind” corrections include (1) instrument temperature fluctuations, (2) long-term instrument drift, (3) the component of the spacecraft velocity along the LOS, (4) the component of Earth rotation along the LOS \cite{niciejewski2006timed}. The main issue may be long-term instrument drift (which is time-dependent) or/and instrument temperature fluctuations (which is dependent on orbit and telescope, since it is dependent on TIMED's orientation towards the Sun). Taking all the possible sources into consideration, in-depth investigations to locate the sources of the differences and calibrate the data are complicated, which is beyond the scope of the paper and will be done in future work.

\begin{figure}
    \noindent\includegraphics[width=10cm]{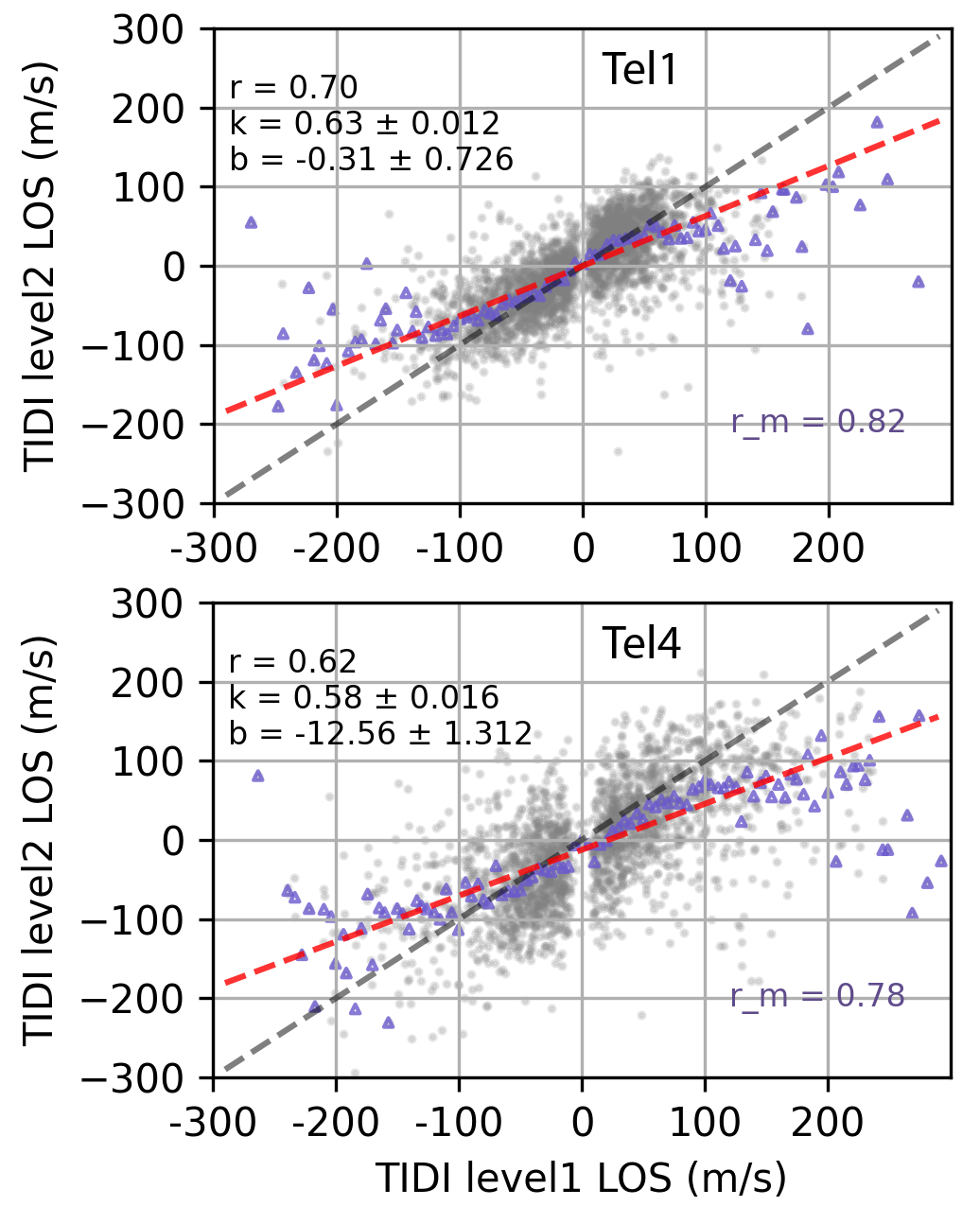}
    \caption{\raggedright Point-to-point comparison between TIDI level 1 and level 2 measurements on January 1, 2020, from 70 km to 120 km for Tel1 (top) and Tel4 (bottom). The red lines represent linear regression fittings to the two datasets and the grey dashed lines are y=x. The blue triangles represent the means of LOS winds in each bin of 5 m/s.}
    \label{fig:level1and2}
    \centering
\end{figure}

Besides, the results indicate that TIDI level 1 LOS winds have a larger range of magnitudes than those from MIGHTI in general. When TIDI measurement magnitudes exceed 100 m/s, more coincident ICON measurements are almost always below 100 m/s. This discrepancy may be due to the “smoothing effects” that are introduced during the inversion of LOS winds. Additionally, TIDI having larger instrument noise would cause more outlier data points above 100 m/s. The purpose of the inversion is to “unsmooth” or “disentangle” the contributions from each altitude in theory. However, in practice, there are a lot of fittings and assumptions applied in the inversion algorithm. For example, for the TIDI level 2 product, the apparent velocity measurements accumulated in a limb scan are inverted by fitting a “scan” of data. Each “scan” of measurements is obtained by moving a single telescope either up or down to view different altitudes during a short time interval. Thus, individual TIDI level 2 LOS profiles are smoother than those of level 1 (Or level 1 data have larger noise than level 2 data). \citeA{harding2017} pointed out that the MIGHTI wind measurements retrieved based on inversion should be interpreted as horizontal and vertical averages and the dominant factor leading to loss of accuracy is horizontal variation of the wind and airglow emission rate, which are smoothed by the long path length of MIGHTI’s line of sight through the atmosphere. For TIDI data products, Figure \ref{fig:level1and2} shows an example of TIDI “raw” (level 1) vs inverted (level 2) LOS winds on January 1, 2020. The coincident measurements of level 1 and level 2 were determined by a window of 2$^\circ$ latitude and 3$^\circ$ longitude. If we discard the measurements with magnitudes $>$ 60 m/s, the slopes of the linear regressions will increase to $>$ 0.9 (very close to the line of y=x). Also, the means (blue triangles) in each velocity bin agree well with the line of y=x when the level 1 LOS winds are below $\sim$60 m/s. This indicates that the two datasets are very consistent with each other for smaller winds; as TIDI level 1 winds become larger ($>$ $\sim$60 m/s), the corresponding level 2 measurements are smaller than those of level 1, but the two datasets are still correlated well with each other in general. Consequently, the inversion algorithm “smooths” the measurements to some degree at least for the TIDI product in practice, though we don't know whether the larger shears in the "raw" LOS winds are real or whether they are instrument noise. \citeA{larsen2002} noted that strong shears exist in the MLT region and the wind maximum between 100-110 km exceeds 100 m/s in more than 60\% of measurements. This is an area of debate, which needs more investigation in the future.

\begin{figure}
    \noindent\includegraphics[width=15cm]{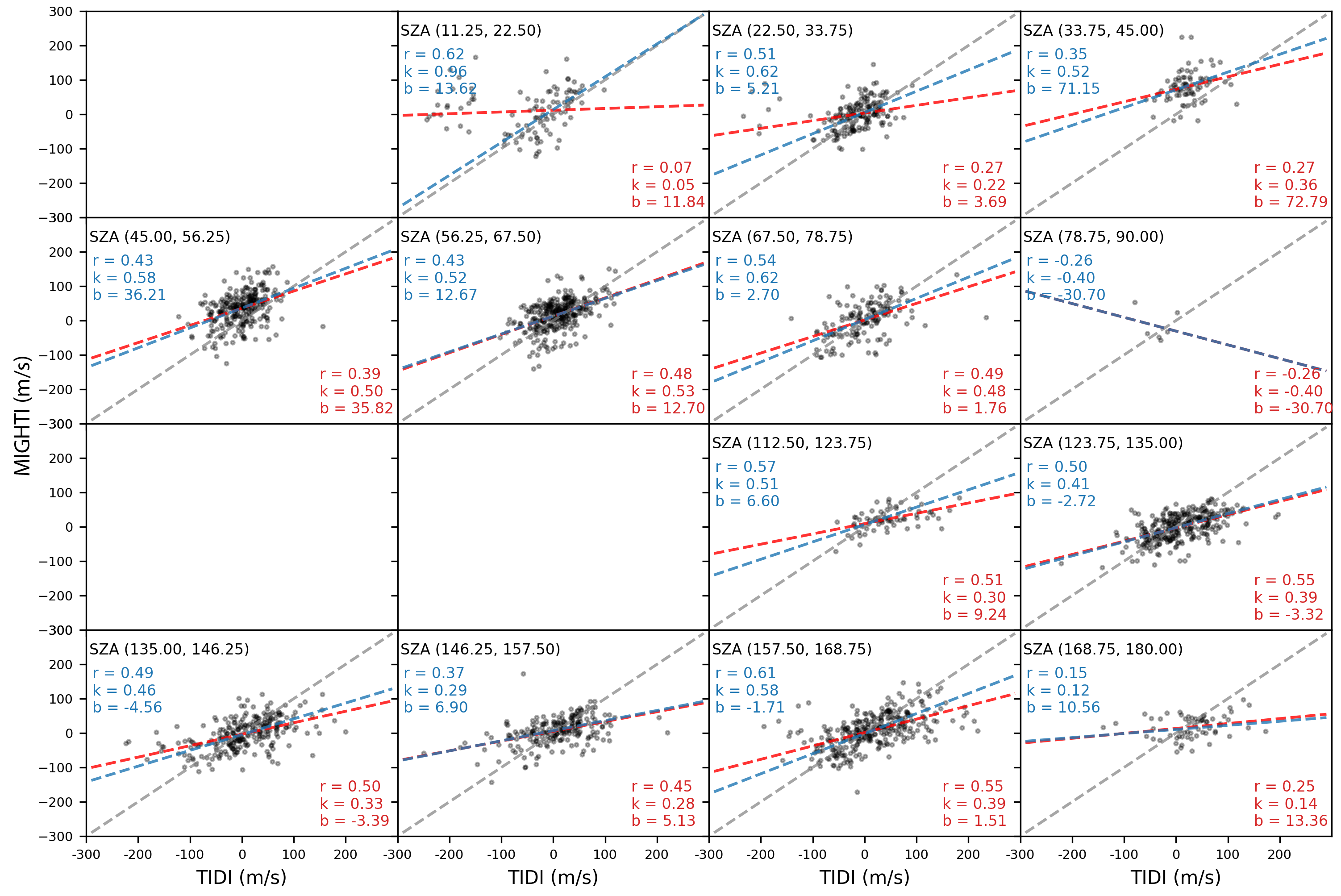}
    \caption{\raggedright Similar to Figure \ref{fig:tel1_sza_merit}, but for Desc/Backward flight. The blue dashed lines represent the regression fittings to the coincidences with TIDI LOS wind magnitudes $<$ 100 m/s and the coefficients are labeled in blue color.}
    \label{fig:tel1_100}
    \centering
\end{figure}

\begin{figure}
    \noindent\includegraphics[width=15cm]{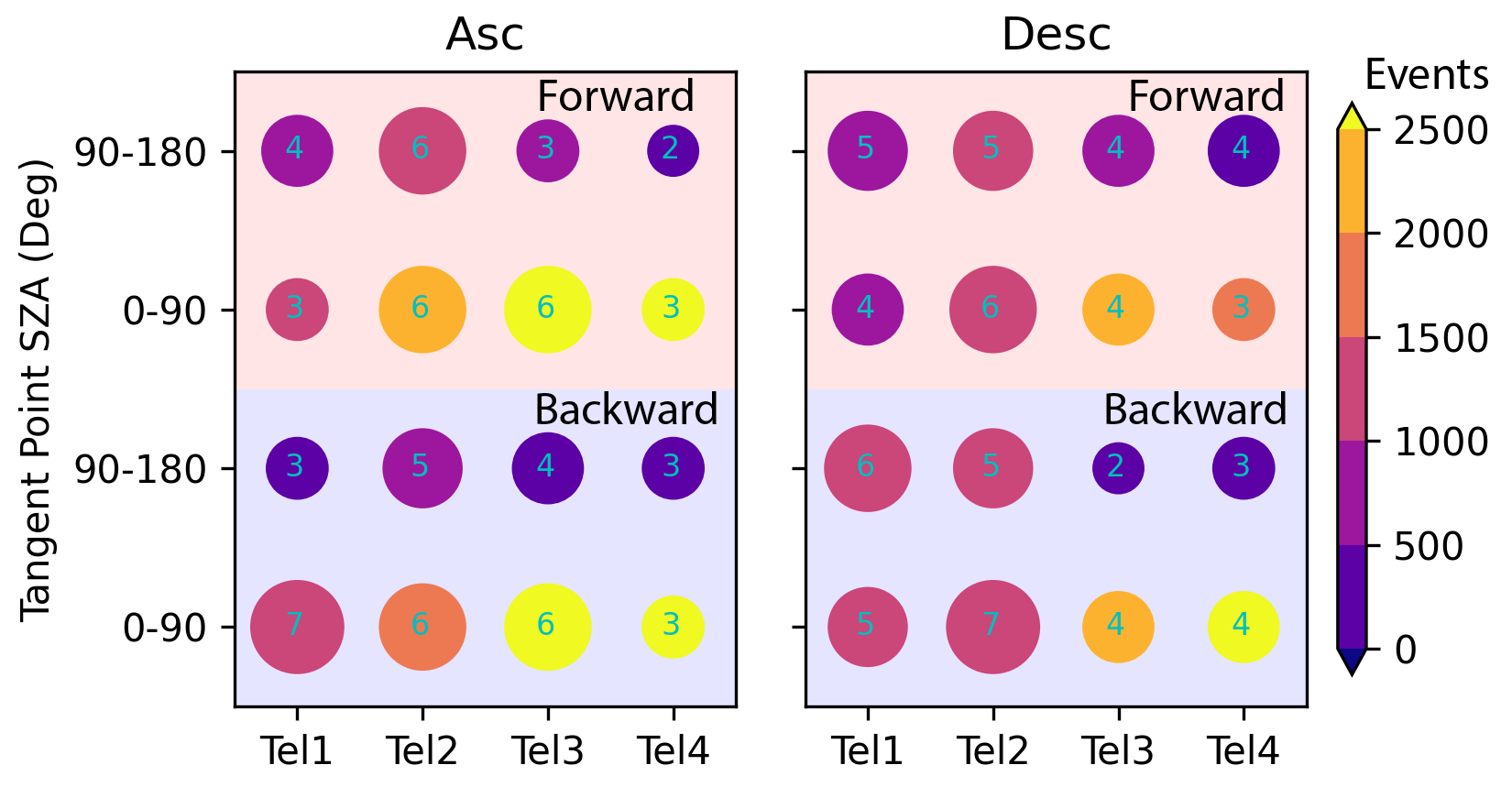}
    \caption{\raggedright Similar to Figure \ref{fig:merits}, but the figures of merit are calculated for the coincidences with TIDI LOS wind magnitudes $<$ 100 m/s.}
    \label{fig:merit_100}
    \centering
\end{figure}

Nevertheless, one can restrict the LOS wind magnitudes to a range of less than 100 m/s to match the two datasets better. Similar plots of TIDI/MIGHTI comparisons were made and figures of merit were calculated for the coincidences with TIDI LOS wind magnitudes $<$ 100 m/s. As an example, Figure \ref{fig:tel1_100} shows the point-to-point comparison of the coincident measurements between Tel1 and MIGHTI for Desc/Backward. When TIDI measurements with magnitudes $>$ 100 m/s are discarded, the regression fittings (blue dashed lines) are improved in some SZA bins, especially from 11.25$^\circ$-33.75$^\circ$ on the dayside. Figure \ref{fig:merit_100} shows the figures of merit with this restriction being applied. The correlation between the two datasets is improved for Tel1/Desc/Day, Tel2/Asc/Day\&Night, and Tel3/Asc/Backward/Day with scores increased by 1. For these measurements, the TIDI LOS winds have relatively high correlations (score $>=$ 5) with MIGHTI LOS winds, and the data with magnitudes larger than 100 m/s may play as outliers or systematic errors (for Tel1). Tel4 scores increase to 4 in the Desc/Backward/Day and Desc/Forward/Night configurations, but the restricted data are still inferior to MIGHTI LOS winds. This is because Tel4 measurements are weakly correlated to MIGHTI observations regardless of magnitudes.

Overall, the figure of merit is provided as a quality flag for TIDI measurements for each satellite configuration and telescope based on the comparisons with MIGHTI observations. For a score $>=$ 5, the TIDI and MIGHTI measurements are consistent. Under this condition in each SZA bin, the correlation coefficient is usually larger than 0.5; the slope tends to be larger than 0.5; and the RMSD between the two datasets is $\sim$50-60 m/s. (2) For a score $<=$ 3, the measurements are very scattered or with systematic errors, showing the weakest correlation. (3) For a score = 4, one should be cautious about the data. Specifically, TIDI data is consistent with MIGHTI wind measurements in the MLT region when certain conditions are met: (1) For Tel1, measurements are consistent with MIGHTI during backward flight (excluding Asc/Backward/Night), and some of those with the ratio of the brightness to standard deviation brightness less than 30 during forward flight; (2) For Tel2, the measurements are generally consistent with MIGHTI across all conditions; (3) For Tel3, consistent measurements are from the Asc/Day configuration; (4) Tel4 data are very scattered showing the weakest correlation with MIGHTI measurements in general; and (5) to match MIGHTI level 2.2 data better, the measurements meeting the requirements above (good data) can be further restricted with the magnitudes less than 100 m/s to exclude outliers. Future work will provide more information on the possible causes of the issues and will attempt to solve them.

\acknowledgments
TIDI level 1 data can be obtained from \url{http://tidi.engin.umich.edu/}. ICON-MIGHTI level 2.2 data are available at \url{https://icon.ssl.berkeley.edu/Data}.
This work is supported by NASA 80NSSC19K0640, 80NSSC21K1086, and 80NSSC21K1890.


%
%
%
%
%

\end{document}